\documentstyle[pre,aps,multicol,epsfig]{revtex}

\def\lsim{\mathrel{\hbox{\rlap{\hbox{\lower4pt\hbox{$\sim$}}}\hbox{$<$}}}}
\def\gsim{\mathrel{\hbox{\rlap{\hbox{\lower4pt\hbox{$\sim$}}}\hbox{$>$}}}}

\begin{document}
\draft
\title{\LARGE \sf{ Evidence concerning Drying behavior of Ne near the Cs surface }}
\author{Francesco Ancilotto$^{1}$, Stefano Curtarolo$^{2}$,
Flavio Toigo$^{1}$, and Milton W. Cole$^{3}$\\}
\address{
$^{1}$Istituto Nazionale per la Fisica della Materia and Dip.to di Fisica G.Galilei \\
Universit\`a di Padova, via Marzolo 8, 35131 Padova, Italy \\
$^{2}$Department of Materials Science and Engineering, MIT, Cambridge, MA 02139, USA\\
$^{3}$Department of Physics, Penn State University, University Park, PA 16802, USA \\
{\rm (\today)}}
\maketitle
\begin{abstract}
Using density functional (DF) and Monte Carlo methods, we have studied the
properties of Ne adsorbed
on a Cs surface, focusing on the region at and near saturated vapor pressure
(SVP).
In the case of Ne/Rb, the experimental data of Hess, Sabatini and Chan are
consistent with the calculations based on an ab initio fluid-substrate potential,
while in the Ne/Cs case there is indication that the potential is $\sim 9\%$
too deep.
In that case, the calculations yield partial drying behavior consistent
with the experimental finding of depressed fluid
density near the surface, above SVP.
However, we find no evidence of a {\it drying transition}, a
result consistent with a mean field calculation of Ebner and Saam.
\end{abstract}

\begin{multicols}{2}
\narrowtext

The subject of wetting has stimulated both basic and applied research
for nearly two centuries. The Young equation implies the possibility of
several alternative varieties of wetting
behavior as a function of the adsorption system and the temperature (T). For example,
at saturated vapor pressure (SVP), a liquid drop may bead up on a surface
exhibiting a finite contact angle $\Theta $. This "non-wetting"
behavior occurs when the following relation between three surface
tensions is satisfied:

\begin{equation}
|\Delta \sigma |\equiv |\sigma _{sv} -\sigma _{sl}| <
\sigma _{lv} \,\,\,\,\,\,\,\,  (0<\Theta <\pi,\,
nonwetting) 
\label{young1}
\end{equation}

Here $v$, $l$, and $s$ refer to vapor, liquid and solid, respectively.
An alternative possibility is "wetting" (sometimes called "complete wetting")

\begin{equation}
\Delta \sigma \ge
\sigma _{lv} \,\,\,\,\,\,\,\,   (\Theta =0, \,wetting) 
\label{young2}
\end{equation}
in which case a very thick liquid film (in equilibrium) is spread uniformly
across the surface. The transition between these regimes
occurs at the wetting temperature $T_W$.
Such a transition, anticipated by general arguments by Cahn \cite{cahn} and Ebner and Saam
\cite{es}), has been seen in a number of instances where the gas-surface
attraction is weak. The theory envisions a third possible scenario, "drying",
in the case of a {\it very} weakly attractive interaction. The criterion is:

\begin{equation}
\Delta \sigma <
-\sigma _{lv} \,\,\,\,\,\,  (\Theta =\pi ,\, drying) 
\label{young3}
\end{equation}

Such behavior is manifested, in principle,
as the presence above SVP of a thick region of vapor
intervening between the surface and the asymptotic bulk liquid.
A recent experiment of Hess, Sabatini and Chan \cite {hess} (HSC)
found some evidence for such drying behavior for 
Ne adsorbed on the Cs surface.
This system is a promising candidate for drying since the
theoretical well depth \cite{chizme} $D\sim 25\, K$ of the gas-surface interaction is
significantly less than the well depth ($\epsilon =33.9\,K$) of the Ne-Ne pair potential.
In this paper, we report calculations relevant to the drying behavior observed
in the HSC experiment.
We have employed two methods, a Density Functional (DF) approach and
grand canonical Monte Carlo (GCMC) simulations;
we have improved somewhat upon techniques used in our previous studies
of wetting transitions \cite{dft1,gcmc}. 
The DF method involves an empirical free-energy functional written
in terms of the density $\rho (\vec{r})$ of the fluid as:

\begin{eqnarray}
F[\rho ] &=&F_{HS}[\rho ]+{\frac{1}{2}}\int \int \rho (\vec{r})\rho (\vec{r}%
^{\,\prime })u_{a}(|\vec{r}-\vec{r}^{\,\prime }|)d\vec{r}d\vec{r}^{\,\prime }
\nonumber \\
&&+\int \rho (\vec{r})V_{s}(\vec{r})d\vec{r} + \gamma F_{id}  \label{energy}
\end{eqnarray}

Here $F_{HS}$ is the free-energy functional for an inhomogeneous hard-sphere
reference system, the second term is the usual mean-field approximation for
the attractive part of the fluid-fluid intermolecular potential $u_{a}$,
while $V_{s}(\vec{r})$ is the external, static adsorption potential due to the
surface. $F_{id} $ is the ideal (classical) gas contribution.
For $F_{HS}$ we use the non-local functional of Ref.\cite{rosinberg}.

The functional (\ref{energy}) involves three parameters:
(i) a correction to the the HS diameter, (ii) a parameter enhancing 
the effective potential well depth of the fluid-fluid interaction, and
(iii) the coefficient $\gamma $ in the last term of (\ref{energy}).
A previous version of the functional (\ref{energy}), involving only
the two parameters (i) and (ii), has been applied with success
to the study of the wetting properties
of Ar and Ne on different surfaces \cite{mist,dft1}.
In particular a remarkably good agreement
has been found \cite{dft1}
between DF calculations for the Ne/Rb system and
the experimental results\cite{hess}, both showing
a first-order wetting transition between 43 and 44 K.
The small revision used here 
is to let the coefficient $\gamma $ in Eq.(\ref{energy}) be a free
parameter; the best fit values are very close, for each value of T,
to the value $\gamma =1$ assumed previously.

\begin{figure}[tbh]
\centerline{\psfig{file=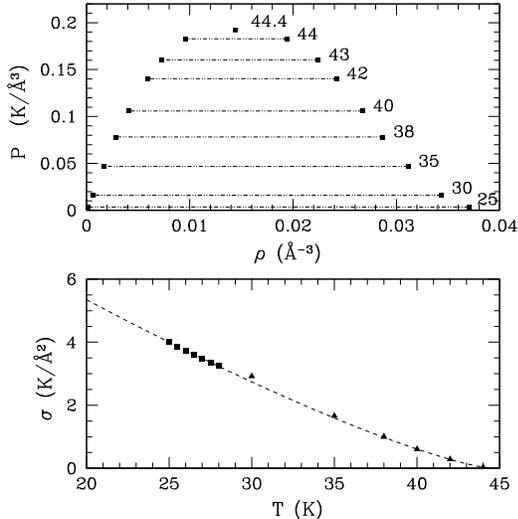,width=68mm}}
\caption{Upper panel: Pressure-density phase diagram 
of bulk Ne at SVP, at various temperatures. The
squares are experimental points at coexistence, 
which are reproduced by construction by our functional.
Lower panel: Liquid-vapor Ne surface tension $\sigma _{lv}$. The squares are experimental points,
the triangles are our calculated values, and the dashed line
is a fit to the experimental data which goes to zero as $(1-T/T_c)^ {\beta }$, with
the exponent $\beta = 1.27$ given by Renormalization Group Theory.}
\label{fig1}
\end{figure}

These three adjustable parameters are
fit to reproduce properties of bulk Ne.
In particular, we require that 
the experimental pressure and densities
for the homogeneous system
are reproduced at liquid-vapor coexistence (see Fig. \ref{fig1}, upper panel), 
and moreover that 
the stability condition $\mu _v = \mu _l$, involving the
chemical potentials of the two phases, is also satisfied.

We then use the functional to study the 
Ne liquid-vapor free interface and 
confirm that it yields good
agreement with the (known) Ne surface tension 
$\sigma _{lv}$ (see Fig. \ref{fig1}, lower panel).

\begin{figure}[tbh]
\centerline{\psfig{file=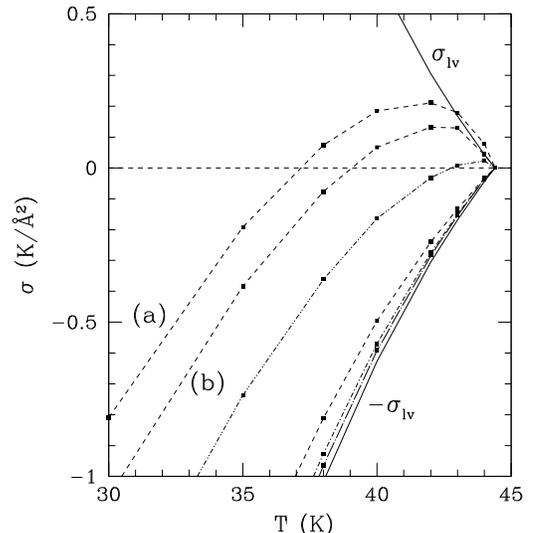,width=68mm}}
\caption{Calculated surface tensions for Ne on different substrates.
The two solid lines show the liquid-vapor surface tension, $\sigma _{lv}$.
Dashed lines labeled (a) and (b): $\Delta \sigma $ for the Ne/Cs system
calculated with the CCZ potential (curve (a)) and the modified-CCZ potential (curve (b)), respectively.
The other lines show $\Delta \sigma $ for Ne on ultraweak 9-3
surfaces, with decreasing values of the potential well depth $D$,
as described in the text.}
\label{fig2}
\end{figure}

\begin{figure}[tbh]
\centerline{\psfig{file=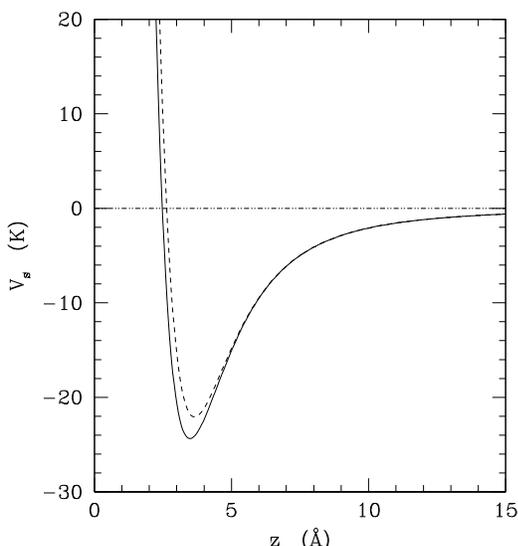,width=68mm}}
\caption{Ne/Cs interaction potentials. The solid line shows
the original CCZ potential, the dashed line shows the modified CCZ potential 
described in the text.}
\label{fig3}
\end{figure}

When studying fluid-surface interactions, it is important to 
employ an accurate adsorption potential. Ab initio potentials
for many adsorption systems were derived recently by Chizmeshya, Cole and Zaremba
\cite {chizme}(CCZ). Overall, these have proven
to be reliable for predicting contact angle and
wetting temperatures of He and Ne on several surfaces \cite{dft1,dft2,bonin}.
Using the Ne/Cs potential, our previous GCMC study \cite{gcmc} found negligible
adsorption of Ne over the entire range ($T < 43$ K) of that study,
that for which the correlation
length $\xi $ of critical fluctuations is smaller than the smallest dimension ($29$\,\AA)
of the periodically replicated simulation cell; this result 
is consistent with the HSC data
for this system.
To more closely approach the critical point ($T_c=44.4\,K$), we have increased
the cell dimensions to $100$\AA$\times100$\AA$\times 200$\AA. Our previous DF study \cite{dft1}
found a wetting transition for Ne/Rb at $T_W\sim 43\,K$,
consistent with the HSC data. With the revised DF method described above 
we confirm the Ne/Rb result and find a {\it wetting}
transition for Ne/Cs at similar temperature ($T_W\sim 43-44\,K$). This is seen in Fig. \ref{fig2} as a
crossing of the two curves $\Delta \sigma (T)$ (curve (a)) and $\sigma _{lv}(T)$
(for the details of the method used to compute the surface tensions $\sigma _{ij}\,(i,j=s,l,v)$, 
see Ref.\cite{dft2}). 
Such a prediction disagrees with the HSC findings for Ne/Cs, of nonwetting at all T. We
conjecture that this discrepancy occurs because the CCZ potential is somewhat too
attractive. We therefore modify the original Ne/Cs potential 
in such a way that 
the modified potential has a slightly smaller well depth than
the original one, as can be seen in Fig. \ref{fig3}.
When using the modified CCZ potential, the wetting transition 
disappears, as can be seen from the curve (b) in Fig. \ref{fig2} which no longer
crosses $\sigma _{lv}(T)$, i.e. 
a rather small ($\sim 9\,\%$) 
correction to the theoretical well depth 
brings the wetting behavior into consistency with experiment.
Such an "error" in the CCZ potential is compatible with the uncertainties
present in its derivation (such as the jellium model of the surface and an
empirical damping procedure applied to the dispersion part of the attraction).

Values of the contact angle $cos(\Theta )\equiv \Delta \sigma /\sigma _{lv}$
can be immediately extracted from Fig. \ref{fig2}. For the modified CCZ
potential we find, for instance, 
$\Theta = 94 ^\circ $ at $T=38\,K$,
$\Theta = 84 ^\circ $ at $T=40\,K$,
and $\Theta = 64^\circ $ at $T=42\,K$.
Measurements of these angles should provide a direct test
of our calculations.

Fig. \ref{fig2} presents results for $\Delta \sigma (T)$
even for quite small hypothetical well depths.
To simulate such ultraweak (UW) adsorbing surfaces, we 
use for simplicity a tunable 3-9
model potential

$$
V_s(z)={4C_3^3 \over 27D^2z^9 }-{C_3    \over z^3 }
$$
where $C_3$ is kept fixed to the Ne/Cs value, while the well depth $D$ is 
arbitrarily varied. 
Fig. \ref{fig2} displays results for the dependence of  $\Delta \sigma (T)$  on the
well depth $D$ for such UW surfaces.
Note that none of these curves crosses the curve $-\sigma _{lv}$.
This means that no drying transition occurs below $T_c$ even for the weakest
interaction considered ($D\sim 0.5 K$), a conclusion that is consistent
with a general (mean field) result \cite{es1} of Ebner and Saam. 
That argument implies that a drying transition
can occur only at $T_c$ as a
consequence of the long-range van der Waals attraction.
In order to check our DF results, we performed extensive GCMC simulations
and confirmed the absence of a drying transition.
What is thus responsible for the apparent drying behavior for Ne/Cs
reported by HSC?
Fig. \ref{fig3} presents a set of density profiles, for the two systems 
investigated in HSC, Ne/Au and Ne/Cs, for different
values of the chemical potential $\mu $ both below and above the 
value $\mu _0$ at liquid-vapor coexistence.
Values of $\mu <\mu _0$ result in profiles that
have the vapor density as the asymptotic 
density $\rho(\infty )$ (far from the surface)
whereas when $\mu > \mu _0$ the liquid density is reached far from the
surface. We note that these DF results
are quite consistent with the GCMC data simulated with the same conditions.
From the profiles shown in Fig. \ref{fig3} one can compute the 
surface {\it excess} coverage per unit area

$$
\Delta N/A \equiv \int _{z_0} ^{\infty } dz [\rho (z)-\rho(\infty )]
$$
(here $A$ is the surface area and the lower limit $z_0$ is taken
at the zero energy turning point of the adsorption potential).

\begin{figure}[tbh]
\centerline{\psfig{file=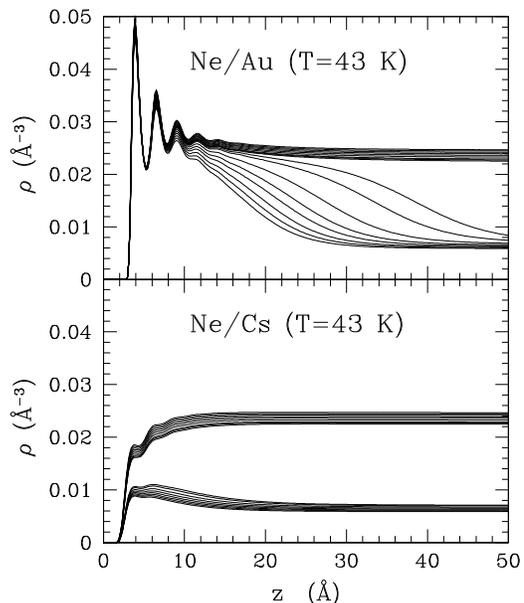,width=68mm}}
\caption{Equilibrium calculated density profiles for Ne/Au and Ne/Cs at $T=43\,K$,
respectively, for different values of the chemical potential
both below and above coexistence. 
$z$ is the coordinate normal to the surface, measured from the surface plane
position.}
\label{fig4}
\end{figure}

Results for $\Delta n \equiv \Delta N/A $ are shown in Fig. \ref{fig4}, both below and above SVP. 
Note that the excess coverage is positive below $\mu _0$
even if the attraction is very weak, as in the case of Ne/Cs.
Above $\mu _0$, the excess coverage for Ne/Cs is negative, 
i.e. there is a deficiency in mass because
the attraction is so weak. 
A positive $\Delta n $ is found instead for Ne/Au, due to the compression 
of the liquid layers close to the surface induced by
the strong adsorption interaction.
The small excess below SVP is in accord with HSC data.
The defect coverage for Ne/Cs above $\mu _0$ 
is $\Delta n \sim 2.2\times 10^{-8}\,
g/cm^2$, corresponding to $\sim 0.9\,ML$.
The expected shift in frequency in a quartz
microbalance experiment \cite{hess_nota} is $\Delta f = (4f^2/nR) \Delta n \sim 4\, Hz$,
i.e. at least one order of magnitude smaller than the
deficit reported by HSC above SVP.
We believe that this
apparent "discrepancy" is misleading. The experiment measures
missing mass by its effect on the resonant frequency shift of the shear
wave of the microbalance. However, a contributing
factor is that a fluid which interacts
weakly with a surface exhibits significant slip,
as reported by Thompson and Robins \cite{robbins}.
Hence the reported mass deficiency implicitly includes
the reduction in mass dragged by the oscillator (as well as the reduced density 
of the neighboring fluid).

Wetting and drying
phenomena in planar geometries have been extensively studied
in the last decade by using MD computer simulations and DF calculations.
In particular, model system of LJ fluids interacting with a
9-3 wall were commonly considered.
A drying transition between the triple-point $T_t$ and 
the critical temperature $T_c$ is invariably
observed in these simulations
for a sufficiently low value of the ratio between the fluid-substrate
and fluid-fluid interaction strengths, although with some controversy concerning
the order of this transition
(see Ref.\cite{henderson,bruin}, and references cited therein).
However, despite the long-range character of the fluid-wall
potential used in these simulations, the interaction 
is usually cut-off, for computational reasons, at
a short distance ($r_c\sim (2-3) \,\sigma _{LJ} $), thus resulting in an
effective {\it finite}
range fluid-wall interaction.
Inclusion of the full range of such interaction would results,
as clearly shown by our results,
in the absence of such drying transition for any $T<T_c$.

\begin{figure}[tbh]
\centerline{\psfig{file=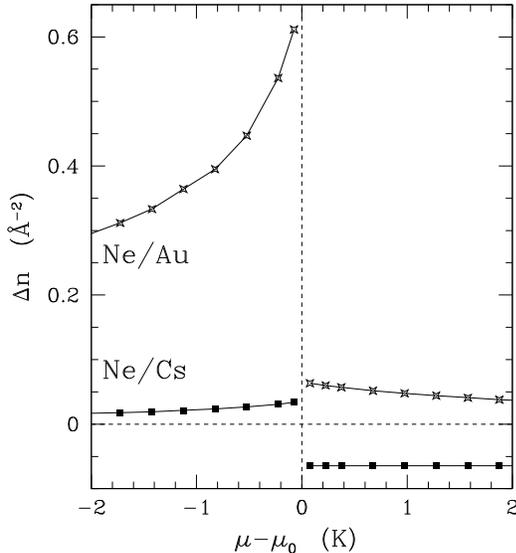,width=68mm}}
\caption{Surface excess coverage $\Delta N/A$ close to coexistence, for Ne/Au and Ne/Cs.
$\mu _0$ is the chemical potential at SVP.}
\label{fig5}
\end{figure}

In summary, DF and GCMC calculations find a wetting
transition for Ne/Cs if the ab initio potential is used to 
describe the fluid-substrate interaction, in disagreement with
experiments. A plausible, small change in the attractive potential well depth changes
the prediction to nonwetting for all T. The resulting behavior of the mass excess
per unit area $\Delta n$ is qualitatively consistent with the
experimental data. Below SVP, $\Delta n$ is small
and positive. Above SVP, an apparent
discrepancy in the magnitude of  $\Delta n$ is attributed to the fact that significant slip
occurs
due to the weak interaction and resulting low density is suggested by
molecular dynamics results of \cite{robbins}.
Finally, our results are consistent with a general prediction of Ebner and Saam that a drying
transition is shifted to $T_c$
by the long-range gas-surface attraction.
Our results (to be reported in detail elsewhere) should be tested by contact angle and
ellipsometric measurements of film thickness for Ne/Cs.

\bigskip
\acknowledgments
We acknowledge helpful discussions with J. Banavar, L. Bruschi,
M. H. Chan, G. Hess, G. Mistura and W. Saam.
This research was supported by the Petroleum Research Fund of the
American Chemical Society.

\end{multicols}

\newpage


\begin{references}
\bibitem{hess_nota} The values entering the prefactor $(4f^2/nR)$ are taken from Ref.\cite{hess}.
\end{references}
\end{document}